\documentclass[final,5p,times,twocolumn]{elsarticle}

\usepackage[utf8]{inputenc} 
\usepackage[T1]{fontenc} 
\usepackage{lmodern} 
\usepackage{microtype} 
\usepackage[english]{babel} 

\usepackage{times} 
\usepackage{amssymb,amsbsy,amsmath,amsfonts,amsthm, yhmath} 
\usepackage{mathtools, cuted}

\usepackage{slashed,bm}
\usepackage{lipsum}

\usepackage{xcolor}

\usepackage{epstopdf}

\usepackage{tabularx}
\usepackage{multirow}
\usepackage{longtable}
\usepackage{threeparttable}

\usepackage{hyperref}
\hypersetup{colorlinks=true, linkcolor=blue, 
citecolor=cyan, urlcolor=red }
\usepackage{url}

\biboptions{sort&compress}

\usepackage{orcidlink}

\allowdisplaybreaks

\journal{Physics Letters B}
\begin{document}

\begin{frontmatter}
\title{Light-quark mass dependence of the $\Lambda(1405)$ resonance
 }
\author[HIM]{Xiu-Lei Ren\orcidlink{0000-0002-5138-7415}}
\affiliation[HIM]{organization={Helmholtz Institute Mainz},
      addressline={Staudingerweg 18}, 
            city={Mainz},
            postcode={55128}, 
            country={Germany}}

\begin{abstract}
We present the light-quark mass dependence of the $\Lambda(1405)$ resonance at leading order in a renormalizable framework of covariant chiral effective field theory. The meson-baryon scattering amplitudes, which are obtained by solving the scattering equation within time-ordered perturbation theory, follow the quark mass trajectory of the Coordinated Lattice Simulations consortium.  At $M_\pi\approx 200$ MeV and $M_K\approx 487$ MeV, our parameter-free prediction of $\Lambda(1405)$ poles is consistent with the recent lattice results of BaSc Collaboration~[Phys. Rev. Lett. 132, 051901 (2024)]. Varying the pion mass from $135$ MeV to $400$ MeV, we present the evolution of double-pole positions of $\Lambda(1405)$: the higher pole remains a resonance around the $\bar{K}N$ threshold; whereas the lower pole undergoes a transition from resonance to a virtual state, and ultimately to a bound state of the $\pi\Sigma$ system, which could be verified by the forthcoming lattice QCD simulations.
\end{abstract}


\end{frontmatter}

\section{Introduction}
Studying the meson-baryon scattering in the SU(3) sector can deepen our understanding of nonperturbative QCD. In particular, the attractive interaction between anti-kaon ($\bar{K}$) and nucleon ($N$), which triggers the existence of the $\Lambda(1405)$ resonance~\cite{ParticleDataGroup:2022pth}, the antikaonic- and multi-antikaonic nuclei/atoms~e.g., in Refs.~\cite{Barnea:2012qa,Curceanu:2019uph}, and the possible antikaonic condensation in neutron star~\cite{Kaplan:1986yq,Li:1997zb}, plays an important role in the strangeness nuclear physics~\cite{Gal:2016boi,Tolos:2020aln}.

The $\Lambda(1405)$ state, which is a  $I(J^P)=0(1/2^{-})$ baryonic resonance of strangeness $S=-1$, was first predicted by Dalitz and Tuan~\cite{Dalitz:1959dn,Dalitz:1960du} and then confirmed by hydrogen bubble chamber experiments in the $\pi\Sigma$ spectrum~\cite{Alston:1961zzd}. $\Lambda(1405)$  does not adhere to the prediction of the constituent quark model~\cite{Isgur:1978xj}, thus categorizing it as an exotic candidate. 
With the development of chiral perturbation theory (ChPT)~\cite{Weinberg:1978kz} and its unitization technique --- chiral unitary approach (CUA)~\cite{Oller:2000ma}, Oller and Mei\ss ner~\cite{Oller:2000fj} found that the $\Lambda(1405)$ state could be composed of two dynamically generated $I=0$ poles, which are located between the $\pi\Sigma$ and $\bar{K}N$ thresholds. This further arouses tremendous efforts in experimental and theoretical research to uncover the nature of the $\Lambda(1405)$ resonance. The detailed discussion and the references can be found in the recent reviews~\cite{Mai:2020ltx,Hyodo:2020czb,Meissner:2020khl}
and the section 83 of Particle Data Group (PDG) review~\cite{ParticleDataGroup:2022pth}. Besides, the most up-to-date experimental researches have reported, e.g.,  the photoproduction cross sections of $K^+\Lambda(1405)$ and $\Lambda(1405)$ lineshape at the BGOOD experiment~\cite{BGOOD:2021sog}; the $K^-p\to \pi^0\Sigma^0$, $\pi^0\Lambda$ cross sections near threshold at DA$\Phi$NE~\cite{Piscicchia:2022wmd}; the $\pi\Sigma$ invariant mass spectra in $K^-$-induced reactions on deuteron by J-PARC E31 Collaboration~\cite{J-PARCE31:2022plu}; the $K^-p$ femtoscopic correlations by ALICE Collaboration~\cite{ALICE:2022yyh}. On the theoretical side, the most recent studies can be found e.g., in Ref.~\cite{Lu:2022hwm} to investigate the $\bar{K}N$ interaction up to next-to-next-to-leading order of CUA; in Ref.~\cite{Sadasivan:2022srs} to perform the uncertainty analysis by including the correlations between the pole parameters within the next-to-leading order CUA; in Ref.~\cite{Cieply:2023saa} to constrain the chirally motivated models with the $\pi\Sigma$ photoproduction mass spectra. 
Note that a consensus has yet to be reached on the one-/two-pole structure of $\Lambda(1405)$. Although the latest PDG review~\cite{ParticleDataGroup:2022pth} has listed $\Lambda(1405)$ as a four-star state and the second pole of $\Lambda(1405)$ as $\Lambda(1380)$ with 2-star rating. 

With the rapid progress of computing techniques, lattice QCD simulation is going to be a promising approach to identifying the $\Lambda(1405)$ state and clarifying the controversy of the one-/two-pole structure. The earlier efforts can be found in e.g. Refs.~\cite{Nemoto:2003ft,Menadue:2011pd,Engel:2013ig,Hall:2014uca,Meinel:2021grq}. Most recently, Baryon Scattering (BaSc) Collaboration has performed a first lattice calculation of the coupled-channel $\pi\Sigma$ and $\bar{K}N$ scattering amplitudes~\cite{BaryonScatteringBaSc:2023zvt,BaryonScatteringBaSc:2023ori}. Using a single $N_f=2+1$ ensemble generated by the Coordinated Lattice Simulations (CLS) consortium, they found two poles of the $\Lambda(1405)$ resonance with the pseudoscalar masses $M_\pi\approx 200$ MeV and $M_K\approx 487$ MeV: the higher pole is a resonance just below the $\bar{K}N$ threshold, and the lower pole is a virtual bound state below the $\pi\Sigma$ threshold. Although the finite-volume correction and the finite-lattice spacing effect are not yet implemented, the BaSc results provide an ideal playground for checking/verifying the predictive power of existing phenomenological models and chiral unitary approaches. 

That is one purpose of the current work: to examine our renormalizable approach for the meson-baryon scattering, which was proposed in Ref.~\cite{Ren:2020wid}. Starting with the covariant chiral effective Lagrangians and employing the time-ordered perturbation theory (TOPT), we define an effective potential as the sum of two-particle irreducible contributions of time-ordered diagrams. The corresponding (coupled-channel) scattering equation $T=V+VGT$ can be derived self-consistently in TOPT. At leading order (LO), we take into account the full off-shell dependence in solving the scattering equation. The renormalized scattering amplitude is obtained by utilizing the subtractive renormalization scheme~\cite{Epelbaum:2020maf}. Note that the subtracted terms are fixed by the chiral symmetry, which is different from the often-used strategies in CUA, where the finite-cutoff or the subtraction constant varies to match the data. In this sense, our framework at LO provides a model-independent prediction for the meson-baryon scattering.

As shown in Ref.~\cite{Ren:2020wid}, our approach has been successfully applied to the pion-nucleon system.
Then, we extend this framework to the strangeness $S=-1$ sector with isospin $I=0$ and study the $\Lambda(1405)$ resonance at the physical point~\cite{Ren:2021yxc}. We solve the scattering equation with the four coupled channels $\pi\Sigma$, $\bar{K}N$, $\eta\Lambda$, and $K\Xi$, and obtain the renormalized $T$-matrix in the isospin limit. Two $\Lambda(1405)$ poles are found on the second Riemann sheet: the lower pole ($1337.7-i79.1$ MeV) is close to the $\pi\Sigma$ threshold, the higher pole ($1430.9-i8.0$ MeV) just below $\bar{K}N$ threshold. Our LO results are consistent with the ones of the next-to-leading order (NLO) study in CUA~\cite{Mai:2014xna}. 

Coming to the unphysical quark mass region, it is interesting to extrapolate our parameter-free prediction of $\Lambda(1405)$ poles and compare it with the BaSc lattice data. If such an agreement is reached, it indicates that our approach exhibits the predictive capability to some extent. Therefore, it enables us to investigate the quark mass dependence of the $\Lambda(1405)$ state. 
Similar studies along this line can be found in Refs.~\cite{Jido:2003cb,Garcia-Recio:2003ejq,Molina:2015uqp,Bruns:2021krp,Xie:2023cej, Guo:2023wes}.
Here we focus on the results along with the quark-mass trajectory of the CLS ensembles. Our model variables, the pseudoscalar meson masses, octet baryon masses, vector meson masses, and meson decay constants, have been calculated in Refs.~\cite{RQCD:2022xux,Ce:2022eix,Ce:2022kxy} based on the CLS configuration. It provides a good opportunity to investigate the light-quark mass dependence of $\Lambda(1405)$, which could be verified by the forthcoming results of the BaSc Collaboration. Note that this LO study can be thought as a first step and the higher order corrections are planned in the future to refine our conclusion.

This article is organized as follows. In Sect.~\ref{SecII}, we briefly present our renormalizable approach to study the $S=-1$ meson-baryon scattering, followed by the description of the CLS lattice data. The comparison with the BaSc results and the quark mass dependence of $\Lambda(1405)$ poles are given in Sect.~\ref{SecIII}. Finally, we summarize our study in Sect.~\ref{SecIV}.

\section{Formalism}~\label{SecII}
In this section, we first lay out the formalism to study the $S=-1$ meson-baryon scattering in the SU(3) unitarized chiral EFT based on a renormalizable framework. The details can be found in Ref.~\cite{Ren:2021yxc}. After that, we investigate the light-quark mass dependence of our model variables by the combination of chiral formulae and lattice QCD data based on the CLS configurations in order to study the evolution of $\Lambda(1405)$ poles. It is worth noting that the three-body effects from $\pi\pi \Lambda$ states could be involved in the $\Lambda(1405)$ region, particularly as we move closer to the physical point. This complicated estimation is beyond the scope of this Letter, and thus we do not include it, as is usually done in the literature. 
Interested readers can refer to e.g. Refs.~\cite{Sadasivan:2021emk, Draper:2023xvu} for more details about the three-body scatterings. 

\subsection{Meson-baryon scattering in TOPT}
At LO, the meson-baryon scattering potential contains the vector-meson exchange contribution, the Born and crossed-Born terms. Note that we employ the vector-meson exchange (VME) to saturate the often-used Weinberg-Tomozawa (WT) term since the former has a better ultraviolet behavior without changing the low-energy physics. 
For the strangeness $S=-1$ sector with isospin $I=0$, the LO interaction for the process $M_i(q_1)+B_i(q_2) \to M_j(q_2)+B_j(p_2)$ is given as in TOPT 
\begin{strip}
\begin{equation}~\label{Eq:LOpotential}
\begin{aligned}
	V_\mathrm{LO} &=  -\frac{1}{32 F_0^2} \sum\limits_{V} C_{M_jB_j,M_iB_i}^V \frac{M_V^2}{\omega_V(\bm{q}_1-\bm{q}_2)} \left(\omega_{M_i}(\bm{q}_1)+\omega_{M_j}(\bm{q}_2)\right) \\
  &\quad \times \Biggl[\frac{1}{E-\omega_{B_i}(\bm{p}_1)-\omega_V(\bm{q}_1-\bm{q}_2)-\omega_{M_j}(\bm{q}_2)}  + \frac{1}{E-\omega_{B_j}(\bm{p}_2)-\omega_V(\bm{q}_1-\bm{q}_2)-\omega_{M_i}(\bm{q}_1)}  \Biggr]\\
& + \frac{1}{4F_0^2}\sum\limits_{B} C_{M_jB_j,M_iB_i}^{B} \frac{m_{B}}{\omega_{B}(\bm{P})} \frac{(\bm{\sigma}\cdot\bm{q}_2)(\bm{\sigma}\cdot\bm{q}_1)}{E-\omega_{B}(\bm{P})} \\
   & + \frac{1}{4F_0^2}\sum\limits_{B} \tilde{C}_{M_jB_j,M_iB_i}^{B} \frac{m_{B}}{\omega_{B}(\bm{K})} \frac{(\bm{\sigma}\cdot\bm{q}_1)(\bm{\sigma}\cdot\bm{q}_2)}{E-\omega_{M_i}(\bm{q}_1)-\omega_{M_j}(\bm{q}_2)-\omega_{B}(\bm{K})} ,
\end{aligned}
\end{equation}
\end{strip}
where $F_0$ and $M_V$ are the pseudoscalar decay constant and the vector meson mass, respectively, in the chiral limit. Both are related via the Kawarabayashi-Suzuki-Riazuddin-Fayyazuddin relation~\cite{Kawarabayashi:1966kd,Riazuddin:1966sw,Djukanovic:2004mm}: $M_V^2=2 g^2 F_0^2$ with the coupling  $g$ of the vector-field self-interaction. The constant coefficients $C^{V}$ of the VME contribution are given in Table~1~of Ref.~\cite{Ren:2021yxc}, and the coefficients of the Born ($C^B$) and crossed-Born ($\tilde{C}^B$) terms, as the functions of the axial vector couplings $D$ and $F$, are given in Tables 2 and 3 of Ref.~\cite{Ren:2021yxc}. In the center-of-mass (c.m.) frame, the four momenta of initial and final states are 
\begin{equation}
\begin{aligned}
q_{1}^\mu &=\left(\omega_{M_i}(\bm{p}), \bm{p}\right),\quad p_{1}^\mu
=\left(\omega_{B_i}(\bm{p}),-\bm{p}\right),\quad  \\
q_{2}^\mu &=\left(\omega_{M_j}(\bm{p}'), \bm{p}'\right),\quad p_{2}^\mu
=\left(\omega_{B_j}(\bm{p}'),-\bm{p}'\right),
\end{aligned}
\end{equation}
with the relative momenta $\bm{p}$, $\bm{p}'$, and the on-shell energy $\omega_X(\bm{p})\equiv \sqrt{m_X^2+\bm{p}^2}$. Furthermore, $E$ is the total energy of the meson-baryon system, and $P=q_1+p_1=q_2+p_2$, $K=p_1-q_2=p_2-q_1$.  Note that the crossed Born term introduces poles when iterated in the scattering equation. To avoid this technical complication, as done in Refs.~\cite{Baru:2019ndr,Ren:2019qow}, we take the total energy $E$ in the denominator as the lowest threshold of the relevant channels. Such approximation does not affect the final results because the contribution of the crossed Born term is rather small to the $S$-wave meson-baryon scattering.

After performing the $S$-wave projection of the LO interaction, one can plug it into the coupled-channel scattering equation in the partial wave basis, 
\begin{equation}\label{Eq:CCVGT}
\begin{aligned}
T^{ji}\left(E; p',p\right) &= V^{ji}\left(E;p',p\right) + \sum_{n}
\int \frac{dk \, k^2}{(2 \pi)^3}V^{jn}(E;p',k)\, \\
&\quad \times  G_n(E) \, T^{ni}(E;k,p),
\end{aligned}
\end{equation}
where $i,~j,~n$ denote the initial, final and intermediate
particle channels of meson and baryon states. The two-body Green function, which is obtained according to the diagrammatic rules of TOPT~\cite{Baru:2019ndr}, has the following form 
\begin{equation}
G_n(E)= \frac{1}{2\, \omega_{M_n}(\bm{k})\, \omega_{B_n}(\bm{k})}\, \frac{m_{B_n}}{E - \omega_{M_n}(\bm{k})-\omega_{B_n}(\bm{k})+i \epsilon} \,.
\label{Gij}
\end{equation}

In order to obtain the renormalized scattering $T$-matrix, it is convenient to separate the leading order potential into the one-baryon reducible part and irreducible part, 
\begin{equation}
  V_\mathrm{LO} = V_{R} + V_{I},	
\end{equation}
where $V_R$ is the Born term, which can separate into two diagrams by cutting only the single-baryon line, and $V_I$ includes the remaining terms, which cannot do such separation. Then, one can rewrite the scattering equation (Eq.~\eqref{Eq:CCVGT}),  $T=V+VGT$, as the three coupled equations: 
\begin{equation}\label{Eq:TandTI_TR}
\begin{aligned}
  T =&\: T_I + (1+T_I\,G)\,T_R\,(1+G\,T_I), \\
  T_I  =&\: V_I + V_I\, G\, T_I, \\
  T_R  =& \: V_R + V_R\, G\,(1+T_I\,G)\,T_R. 	
\end{aligned}
\end{equation}
It is found that the irreducible part $T_I$ is finite in the removed regulator limit. To demonstrate this fact, we take the vector-meson-exchange potential as an example. First, we investigate the UV behavior of the once-iterated contribution $VGV$ at one-loop level: 
\begin{equation}
	I_{VGV} = \int\frac{d^3 k}{(2\pi)^3} V_\mathrm{VME}(p',k) \, G(k) \, V_\mathrm{VME}(k,p)  ,
\end{equation}
When the integral momentum $k$ goes to infinity, the one-loop integral becomes
\begin{equation}
	I_{VGV} \to \int \frac{d^3 k}{(2\pi)^3} \frac{1}{k}\, \frac{1}{k^3}\, \frac{1}{k}, 
\end{equation}
which converges. This benefit arises from the VME potential, which has a milder UV behavior compared to the often-used WT term. One can easily verify that the infinite iteration of $V_\mathrm{VME}$ in the ladder diagram also converges. Besides, the crossed-Born term presents a similar UV behavior to the VME counterpart. Because the half-off-shell potential $V_\mathrm{CB}(k,p)$ approaches 
\begin{equation}
	V_\mathrm{CB}(k,p) \to  \frac{\bm{\sigma\cdot \bm{p}} \, \bm{\sigma}\cdot{\bm{k}}}{k^2},
\end{equation} 
when the integral momentum $k$ goes to infinity. 

Regarding the reducible part, because the iteration of the Born term is quadratically divergent, thus the only divergent term in the total $T$-matrix originates from the reducible part  $T_R$. To achieve renormalization, we first rewrite the Born potential in a separable form, 
\begin{equation}
	V_{R}(p',p;E) = \xi^{T}(p')\, C(E)\, \xi(p), 
\end{equation}
where $\xi^T(p)$ is defined as  $\xi^T(p)=(1,\, p)$, and $C(E)$ denotes a $2\times 2$ matrix. Then, the reducible $T_R$ matrix can be expressed in a separable form 
\begin{equation}
	T_{R}(p^{\prime},p;E)=\xi^{T}(p^{\prime})\, \chi(E) \, \xi(p),
\end{equation}
where $\chi(E)$, as given below, is divergent. 
\begin{equation}
	\chi(E)=\left[C(E)^{-1}-\xi \, G(E)\, \xi^{T}-\xi \, G(E)\, T_{I}^{S}\, G(E) \, \xi^{T}\right]^{-1}.
\end{equation}
Then, one can apply the subtractive renormalization~\cite{Epelbaum:2020maf} by replacing the meson-baryon Green function $G(E)$ with the subtracted one
\begin{equation}
G^S(E)=G(E)-G(m_{B}),
\end{equation}
to systematically remove the divergences. This corresponds to taking into account the counter-terms generated by the renormalization of the baryon masses and the meson-baryon coupling constants. Note that such subtraction, which is an analogy to the extended-on-mass-shell (EOMS) scheme~\cite{Fuchs:2003qc}, corresponds to the expansion around the threshold and can systematically remove the chiral power-counting breaking terms in the iterated amplitudes. Finally, we obtain a renormalized $T$-matrix
\begin{equation}
T = T_{I}^S+\left(\xi^{T}+T_{I}^S\, G^{S}\, \xi^{T}\right) \, \chi^{S}(E) \, \left(\xi+\xi\, G^{S}\, T_{I}^S \right),
\end{equation}
where the superscript $S$ demonstrates the replacement of $G$ as $G^S$ in the expressions of $T_I$ and $\chi(E)$.

\subsection{Quark mass dependence of our model variables}
First, we investigate the quark mass dependence of the pseudoscalar meson masses $M_P$ and the octet baryon masses $m_B$, which appear in both of the LO chiral potential and the two-body Green function. 
Confronting the recent lattice results of  $\Lambda(1405)$~\cite{BaryonScatteringBaSc:2023zvt,BaryonScatteringBaSc:2023ori}, we conduct the chiral analysis of the meson and baryon masses from the RQCD Collaboration~\cite{RQCD:2022xux} based on the CLS configuration, which is also utilized by the BaSc Collaboration.  

\begin{figure}[t]
    \centering
    \includegraphics[width=0.48\linewidth]{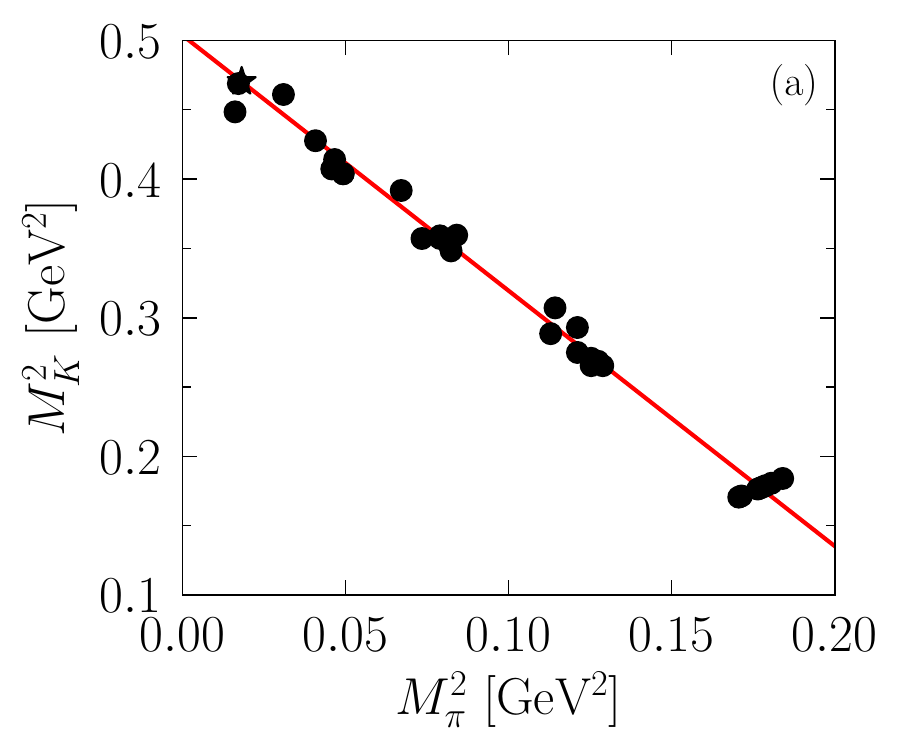} ~
    \includegraphics[width=0.48\linewidth]{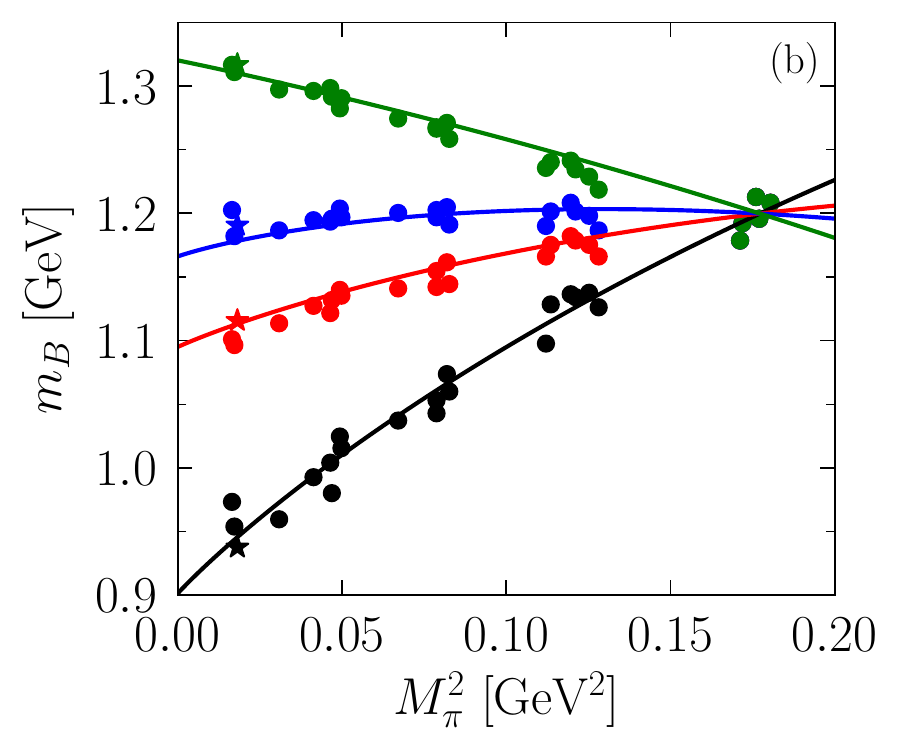}
    \caption{Left panel: Linear dependence of $M_K^2$ in terms of $M_\pi^2$ for the quark mass trajectory $\overline{m}=m_\mathrm{symm}$. Right panel: The octet-baryon masses as the functions of the pion mass with the mass trajectory $\overline{m}=m_\mathrm{symm}$. The dots are the lattice results, the stars stand for the experimental masses, corrected for the isospin-breaking effects.}
    \label{Fig:MpiVSMKandmB}
\end{figure}

First, we use a linear function of $M_\pi$ and $M_K$
\begin{equation}
  M_K^2 = a + b M_\pi^2,	
\end{equation}
as done in Ref.~\cite{Ren:2012aj}, 
to parameterize the quark mass dependence of the kaon mass for the quark mass trajectory of $\overline{m}=m_\mathrm{symm}$, keeping the trace of the bare quark-mass matrix fixed $Tr(M) = \mathrm{const.}$, given by the CLS ensembles. The lattice pion and kaon masses in Ref.~\cite{RQCD:2022xux} have been corrected by including the finite-volume effects. The values of the pion and kaon masses at the physical point are 
\begin{equation}
	M_{\pi,\text{phys}}=134.8(3)~\mathrm{MeV},\quad M_{K,\text{phys}}=494.2(3)~\mathrm{MeV}, 
\end{equation}
 representing the electrically neutral isospin-averaged results~\cite{Aoki:2016frl}. The best fitting result is shown in the left panel of  Fig.~\ref{Fig:MpiVSMKandmB} with $a=0.25189$~GeV$^{2}$ and $b=-0.42143$. 
Then, the quark mass dependence of $M_\eta$ is determined via the Gell-Mann-Okubo relation $ M_{\eta}^2=(4 M_{K}^2 - M_{\pi}^2)/3$. 

As to the lowest-lying octet baryon masses, we employ the mass expressions at next-to-leading order in covariant baryon ChPT
\begin{equation}
\begin{aligned}
	m_B &= m_0 + \sum\limits_{P=\pi,~K}\xi_{PB}^{(2)}M_P^2 \\
	&\quad + \frac{1}{(4\pi F_0)^2}\sum\limits_{P=\pi,~K,\eta}\xi_{PB}^{(3)}H_B\left(\frac{M_P}{m_0}\right) ,
\end{aligned}	
\end{equation}
where $m_0$ is the baryon mass in the chiral limit. The LO coefficients $\xi_{PB}^{(2)}$, as the function of low-energy constants (LECs) $b_{0,~D,~F}$, can be found in Table 1 of Ref.~\cite{Ren:2012aj}. The NLO coefficients $\xi_{PB}^{(3)}$, the combination of the baryon axial coupling constants $D$ and $F$, are given in Table 2 of Ref.~\cite{Ren:2012aj} with the one-loop function $H_B(x)$
\begin{equation}
	H_B(x) = -2x^3\Biggl[ \frac{x}{2} \log(x) +  \sqrt{1-\frac{x^2}{4}}\arccos\left(\frac{x}{2}\right)\Biggr],
\end{equation}
which is obtained in the EOMS scheme~\cite{Fuchs:2003qc}. 
The $F_0$ is the pseudoscalar meson decay constant in the chiral limit. To obtain the light-quark mass dependence of the octet baryon mass, we perform the fit of the RQCD results~\cite{RQCD:2022xux}, where the finite-volume corrections and the discretization effects are taken into account. Furthermore, we also include the experimental value of octet baryon masses, where the isospin-breaking effects from QED and QCD are removed~\cite{RQCD:2022xux}. We use the SU(6) relation  $F=2/3D$ and $D+F=g_A=1.267$ to fix $D=0.760$ and $F=0.507$. For the meson decay constant, we use the SU(3) averaged value $F_0=1.17 F_\pi$ with $F_\pi=92.4$ MeV to achieve a relatively better description of lattice data. Then, we have four free parameters, and the best-fit results lead to $m_0=0.813$ GeV, $b_0=-0.669$ GeV$^{-1}$, $b_D=0.0254$ GeV$^{-1}$, and  $b_F=-0.399$  GeV$^{-1}$. The light-quark mass dependence of the baryon masses  is shown in the right panel of Fig.~\ref{Fig:MpiVSMKandmB}. 

\begin{figure}[t]
    \centering
    \includegraphics[width=0.49\linewidth]{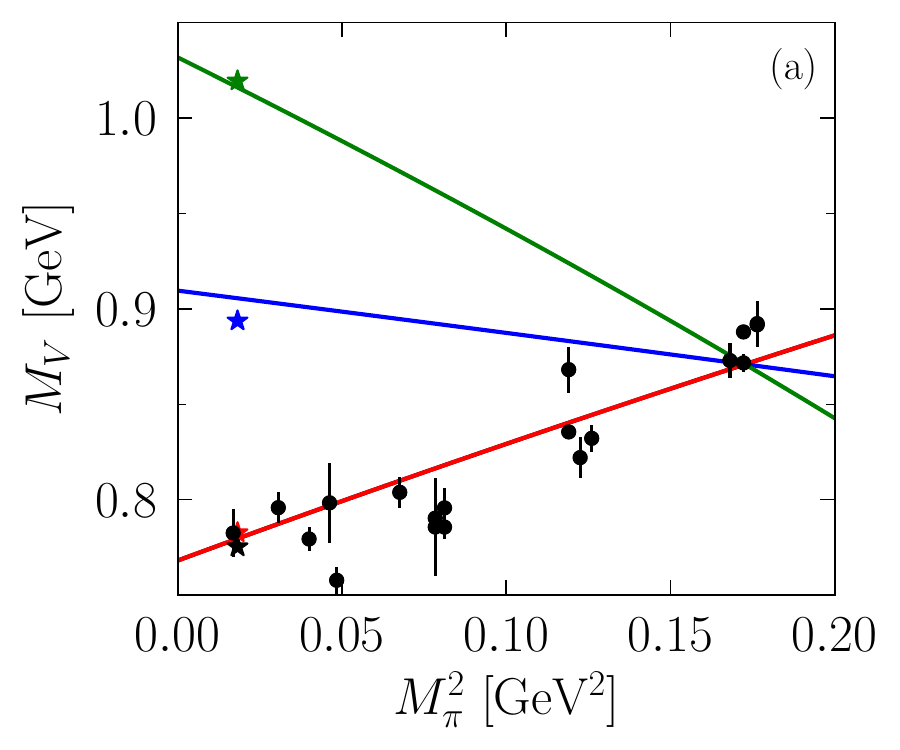}~
    \includegraphics[width=0.49\linewidth]{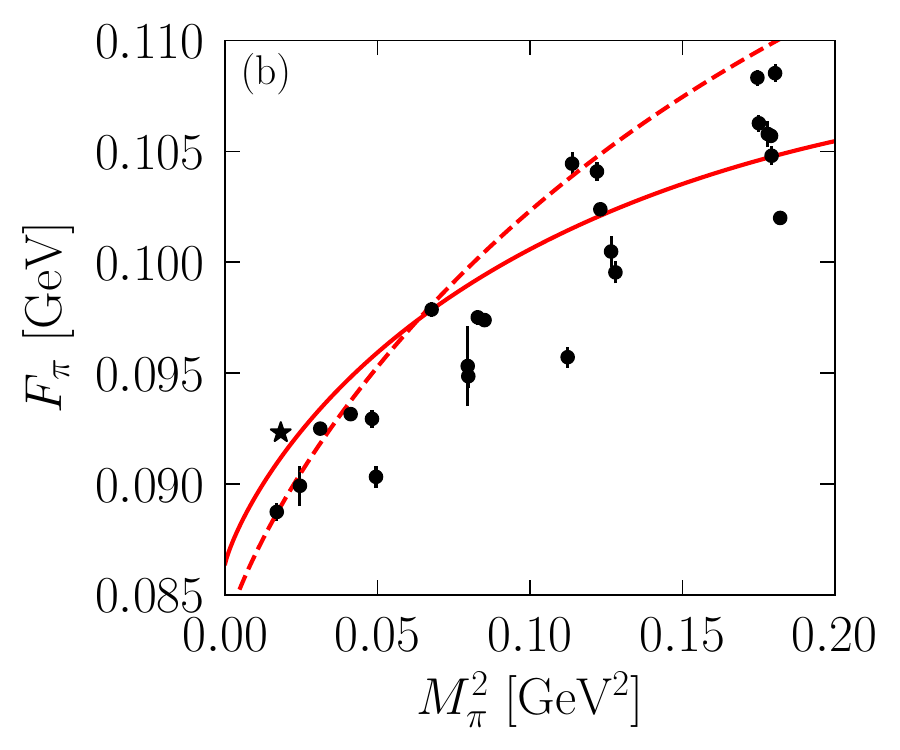}
    \caption{The vector meson masses (left panel) and the pion decay constant (right panel) as the functions of the pion mass with the quark mass trajectory $\overline{m}=m_\mathrm{symm}$. Notations are the same as Fig.~\ref{Fig:MpiVSMKandmB}.}
    \label{Fig:MpiMVandFpi}
\end{figure}

Besides, we also require the vector meson mass $M_V$ and the decay constant $F_0$ in the chiral limit to obtain the LO potential~(Eq.~\eqref{Eq:LOpotential}). In practice, we use the experimental masses of vector mesons $M_{\rho,~\omega,~K^*,~\phi}$ and the pion decay constant $F_
\pi$ or the SU(3) averaged decay constant in Ref.~\cite{Ren:2021yxc}. The introduced difference between the chiral limit values and the physical values of the the vector meson masses and the meson decay constant is of higher order. As to the light-quark mass dependence of $M_V$, only rho meson mass $M_\rho$ is simulated with the quark mass trajectory $\overline{m}=m_\mathrm{symm}$ of the CLS configuration by Mainz lattice group~\cite{Ce:2022eix}. To determine the quark mass dependence of the vector meson masses, we use the results of LO ChPT in the SU(3) sector~\cite{Zhou:2014ila},
\begin{equation}\label{Eq:VectorMassLO}
\begin{aligned}
M_\rho^2  &= M_\omega^2  = M_0^2+2 \lambda_m M_\pi^2+\lambda_0\left(2 M_K^2+M_\pi^2\right), \\
M_{K^*}^2 & =M_0^2+2 \lambda_m M_K^2+\lambda_0\left(2 M_K^2+M_\pi^2\right), \\
M_\phi^2 & =M_0^2+4 \lambda_m M_K^2-2 \lambda_m M_\pi^2+\lambda_0\left(2 M_K^2+M_\pi^2\right), 
\end{aligned}
\end{equation}
where $M_\rho=M_\omega$. The three LECs $M_0$, $\lambda_0$, and $\lambda_m$ are determined by fitting the physical values of vector meson masses and the lattice data of $M_\rho$, resulting in $M_0=0.538$ GeV,  $\lambda_m=0.468$,  $\lambda_0 =0.593$. The reasonable fitting result is presented in the left panel of Fig.~\ref{Fig:MpiMVandFpi}. 

\begin{table*}[t]
\caption{Pole positions of two $\Lambda(1405)$ poles and the corresponding couplings $g_i$ from the $\pi\Sigma$-$\bar{K}N$ coupled channel calculation at $M_\pi=203.7$ MeV and $M_K=486.4$ MeV. Pole positions from the full four-coupled channel calculation are also given. For comparison, the BaSc results are listed with one combined uncertainty.}
\label{Tab:Mpi200Results}
\begin{threeparttable}
  \begin{tabular}{c|c|c|c|c|c|c}
     \hline\hline 
                      & BaSc~\cite{BaryonScatteringBaSc:2023zvt,BaryonScatteringBaSc:2023ori} &  \multicolumn{4}{c}{This work    }\\
                          \cline{2-7}    
      $\Lambda(1405)$   & $z_R$ [MeV]  & $z_R^\dagger$ [MeV] & $z_R$ [MeV]   & $g_{\pi\Sigma}$ &   $g_{\bar{K}N}$  & $|g_{\pi\Sigma}|/|g_{\bar{K}N}|$ \\
    \hline
   Lower pole  & $1392(18)$ &  $1389.05$ &  $1387.14$    & $0.021 + i 1.87$ & $0.017 + i 1.55$  &  $1.21$\\
   Higher pole & $1455(21)-i11.5(6.0)$ &  $1464.55 - i 9.44$ &  $1469.86 - i 4.71$ & $0.038 + i 0.98$  & $1.51 - i 1.22$  & $0.50$ \\
   \hline\hline 
  \end{tabular}
  \begin{tablenotes}\footnotesize
\item[$^\dagger$] The full calculation with four coupled channels: $\pi \Sigma$, $\bar{K} N$, $\eta \Lambda$, and $K \Xi$.
\end{tablenotes}
  \end{threeparttable}
\end{table*}

As to the quark mass dependence of the pion decay constant $F_{\pi}$, the lattice results of the CLS configuration are reported in Ref.~\cite{Ce:2022kxy} with the quark mass trajectory $\overline{m}=m_\mathrm{symm}$. From Fig.~\ref{Fig:MpiMVandFpi}, it seems that the lattice data for the small pion masses are slightly lower than the empirical value. 
To describe the lattice results and obtain the pion mass dependence of $F_\pi$, we follow the fitting procedure outlined in Ref.~\cite{Molina:2015uqp}. For further details and formalism, one can refer to the appendix of Ref.~\cite{Molina:2015uqp}. The lowest order $M_{0\pi}$, $M_{0K}$, and the chiral limit value $F_0$ are adjusted by reproducing the experimental value and each set of lattice results of $M_\pi$, $M_K$ and $F_\pi$ given in Ref.~\cite{Ce:2022kxy}. We take two strategies to perform the fit: first, the LECs $L_i^r,~i=4,5,6,8$ are fixed as Fit I of Ref.~\cite{Nebreda:2010wv}, where the experimental and lattice meson-baryon scattering data are fitted. The pion mass dependence of $F_\pi$ is presented as the dashed line in Fig.~\ref{Fig:MpiMVandFpi}-(b). Although the lattice data with small pion masses are well covered, the experimental $F_\pi$ is underestimated and the lattice data with large pion masses are overestimated. To achieve a relatively compatible description, we release $L_{5}^r$ and $L_8^r$ in the fit, which leads to the solid line in Fig.~\ref{Fig:MpiMVandFpi}-(b). In this case, the values of $L_{5,8}^r$ come to $L_5^r=0.484\times 10^{3}$ MeV$^2$ and $L_8^r=1.143\times 10^{3}$ MeV$^2$. Note that $L_8^r$ is slightly changed, but  $L_5^r$ decreases around $1/3$ times in comparison with the value given in Fit I of Ref.~\cite{Nebreda:2010wv}. This observation could call for a high-statistics simulation of $F_\pi$ at small pion masses.

\section{Results and discussion}~\label{SecIII}
We first present our prediction of the $\Lambda(1405)$ poles with unphysical quark masses $M_\pi\approx 200$ MeV and $M_K\approx 487$ MeV. For comparison with the lattice result of BaSc Collaboration~\cite{BaryonScatteringBaSc:2023zvt,BaryonScatteringBaSc:2023ori}, we focus on the $\pi\Sigma$-$\bar{K}N$ coupled channel and use the same hadron masses: $M_\pi=203.7$ MeV, $M_K=486.4$ MeV, $m_N=979.8$ MeV, and $m_\Sigma=1193.9$ MeV to calculate the scattering amplitudes. The vector meson masses appearing in the LO potential are estimated through the chiral expansion~Eq.~\eqref{Eq:VectorMassLO} with $M_\pi=203.7$ MeV. For the decay constant, we take $F_0=F_\pi=93.2$ MeV, as used by the BaSc lattice simulation, instead of the chiral limit value. The difference would be of the higher order. The axial vector couplings are fixed as $D=0.760$ and $F=0.507$ with $D+F=g_A=1.267$. 

To obtain the meson-baryon scattering amplitudes, we first solve the integral equation by including the full off-shell dependence of the kernel potential. Such treatment is always employed in the study of the nucleon-nucleon interaction within the chiral EFT, e.g.~Refs.~\cite{Epelbaum:2008ga,Machleidt:2011zz}. Furthermore, the pole positions of resonances can change in solving the meson-baryon integral loop with the full off-shell dependence~\cite{Bruns:2010sv,Mai:2012dt,Morimatsu:2019wvk} in comparison with the approximation of the on-shell factorization. Secondly, as mentioned above, we employ the subtractive renormalization to obtain the finite $T$-matrix by taking the cutoff limit $\Lambda \to \infty$. In practice, a sharp cutoff regulator $F_\Lambda(p',p)=\theta(\Lambda-p')\,\theta(\Lambda-p)$ is employed with $\Lambda= 10$ GeV, which is sufficiently large to get rid of the finite-artifacts (as shown in the following). Therefore, our framework does not have the usual obstacle of the cutoff dependence of the $T$-matrix present in the traditional chiral unitary approaches, which results in parameter-free predictions.   

We are in the position to search poles appearing in the $\pi\Sigma$-$\bar{K}N$ coupled-channel $T$-matrix. One can perform an analytic continuation of the $T$-matrix into the complex $s$-plane. The 4 Riemann sheets, which are labeled by the sign of the imaginary part of the c.m. momentum $p_i=\lambda^{1/2}(s,M_P^2, m_B^2)/(2\sqrt{s})$ for each channel, are introduced. The most relevant Riemann sheets in the current work are the physical sheet $(+,+)$ and the second Riemann sheet $(-,+)$ with only the $\pi \Sigma$ channel open for decay. Furthermore, in order to investigate the structures of the poles, one can parameterize the on-shell scattering $T$-matrix around the pole position $z_R$,
\begin{equation}
	 T_{ij}\simeq 4\pi \frac{g_i\, g_j}{z-z_R},
\end{equation} 
where the (complex) $g_i$ ($g_j$) couplings are introduced to denote the coupling strength of the resonance pole to the initial (final) transition channel. Those couplings can be extracted from the residues of the $T$-matrix. 

\begin{figure}[b!]
\includegraphics[width=0.45\textwidth]{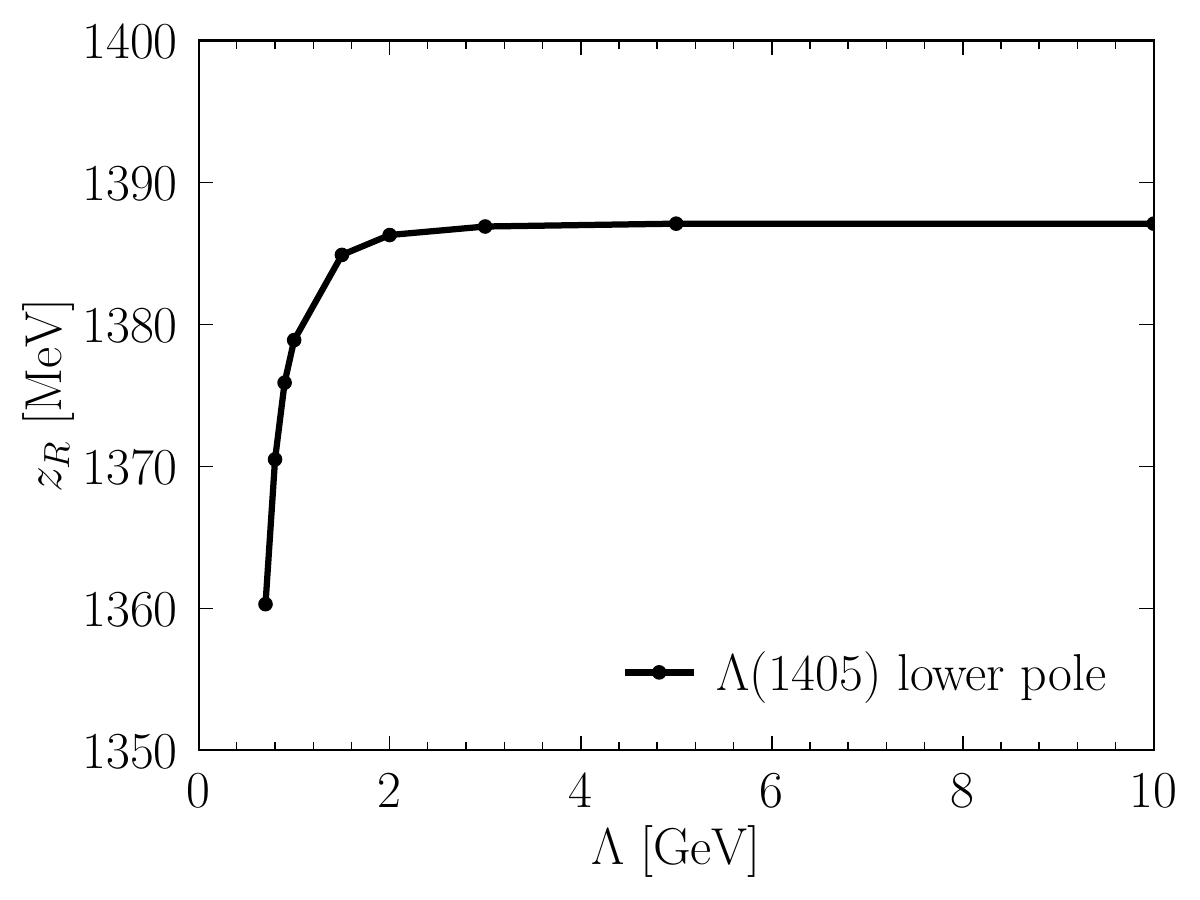}
  \caption{Evolution of lower pole position as a function of the momentum cutoff $\Lambda$. }
 \label{Fig:lowpole_cutoff}
\end{figure}

\begin{figure*}[t]
  \includegraphics[width=0.45\textwidth]{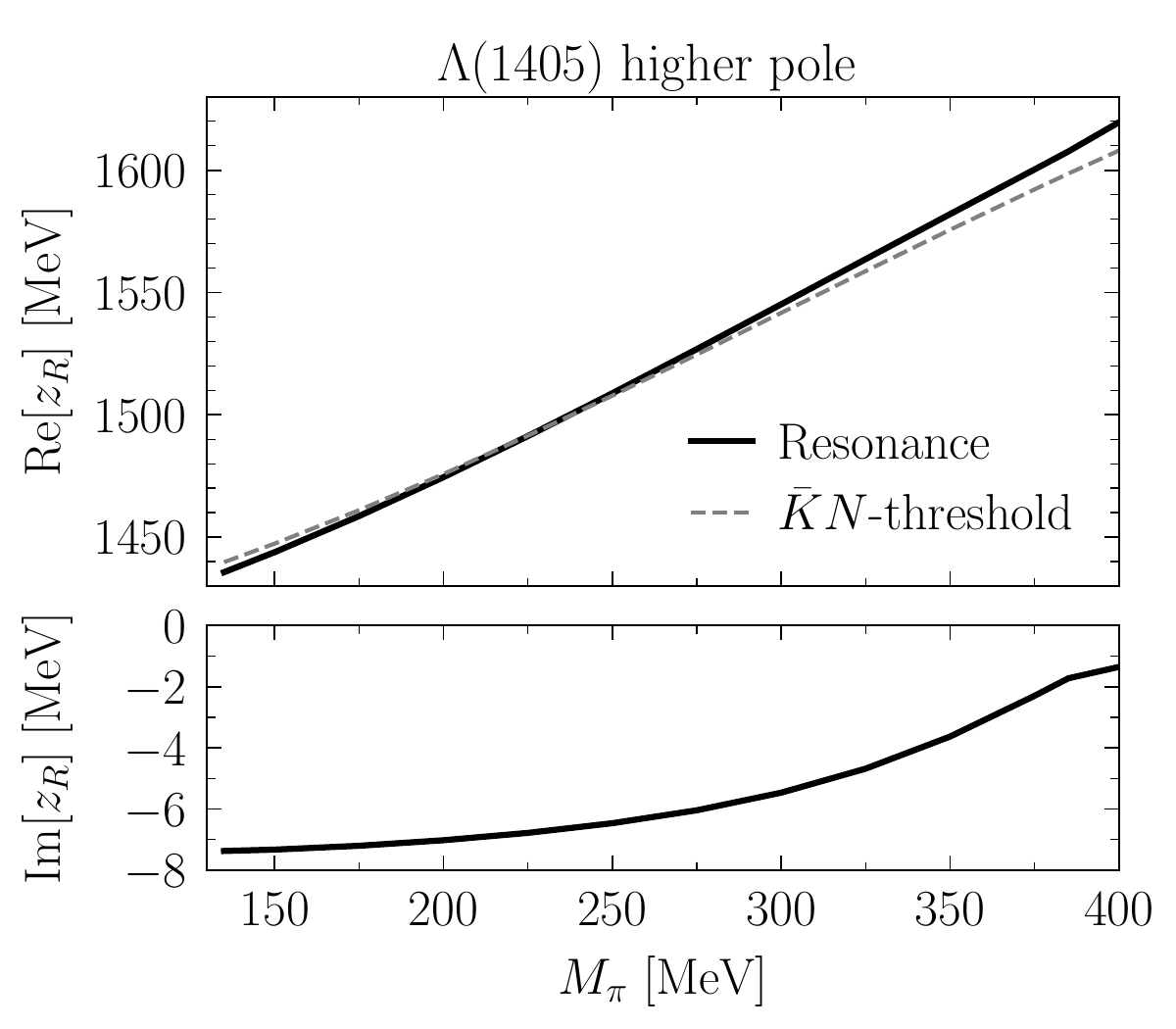}~~
   \includegraphics[width=0.45\textwidth]{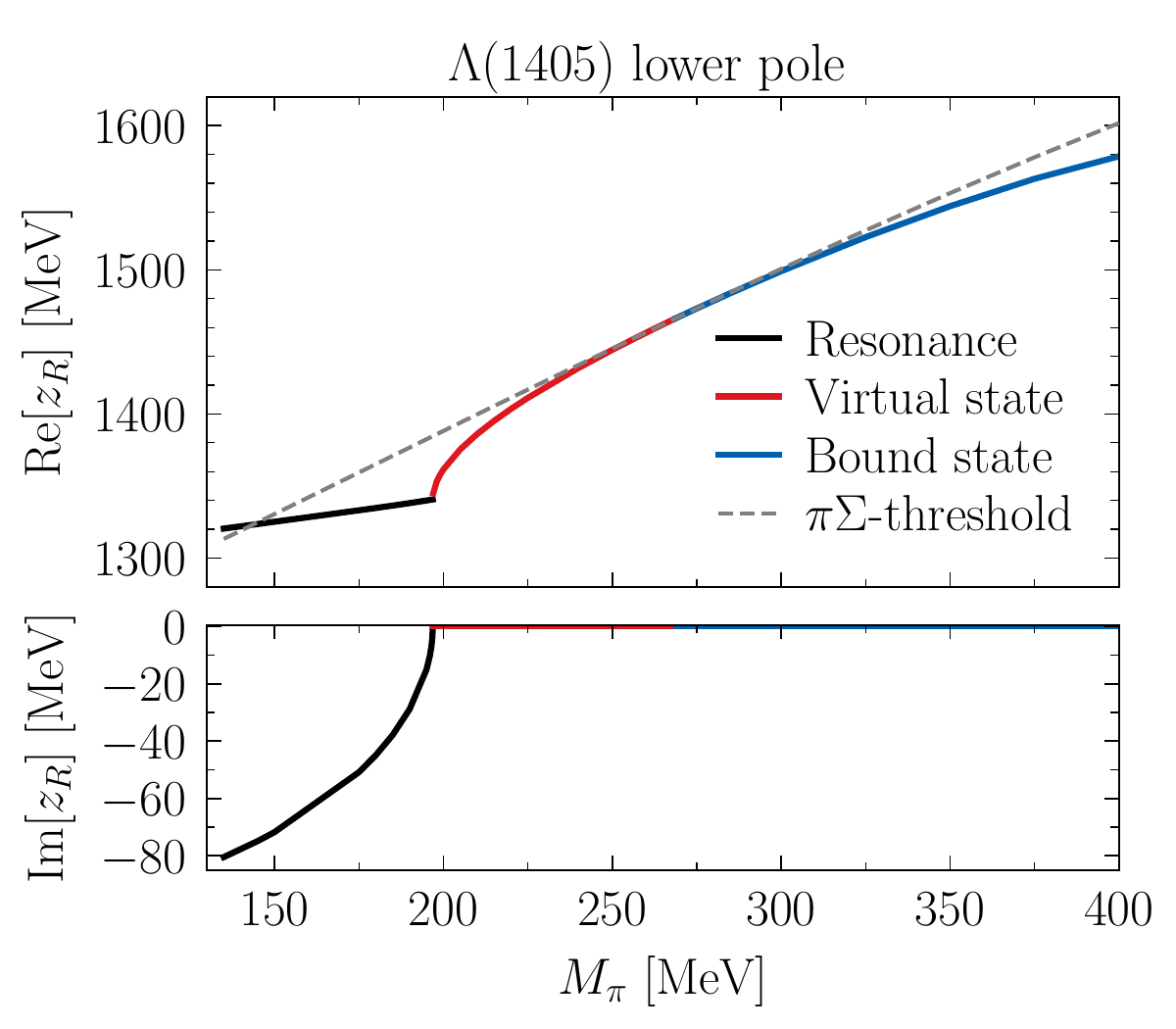}
  \caption{$\Lambda(1405)$ poles as the function of pion mass with the CLS quark mass trajectory $\overline{m}=m_\mathrm{symm}$. The upper (lower) panel is the real (imaginary) part of the pole position. The black-, red-, and blue-solid curves denote the pole as the resonance, virtual state, and bound state, respectively. The dashed lines correspond to the $\bar{K}N$ (left panel) and $\pi\Sigma$ (right panel) thresholds.}
  \label{Fig:Poles_Mpi}
\end{figure*}

We find two poles on the $(-,+)$ Riemann sheet, one is resonance around $\bar{K}N$ threshold (denoted as ``higher pole''), and another is the virtual bound state just below $\pi\Sigma$ threshold (denoted as ``lower pole''). Their pole positions and the coupling strengths are tabulated in Table~\ref{Tab:Mpi200Results}. One can see that our prediction of pole positions is consistent with the lattice results of BaSc Collaboration~\cite{BaryonScatteringBaSc:2023zvt,BaryonScatteringBaSc:2023ori}, except for the width of the higher pole, which  approaches the lower bound of lattice value. 
It is interesting to note that the pole positions ($z_R=1389.05$ MeV and $1464.55 - i9.44$ MeV) are closer to the lattice results if we perform a full calculation with four coupled channels $\pi \Sigma$, $\bar{K} N$, $\eta \Lambda$, and $K \Xi$.  Furthermore, the values of $g_{\pi\Sigma,\bar{K}N}$ and their ratios indicate that the lower pole couples predominantly to the $\pi\Sigma$ channel, while the higher pole couples strongly to the $\bar{K}N$ channel. The ratios of couplings are also consistent with the BaSc results~\cite{BaryonScatteringBaSc:2023zvt,BaryonScatteringBaSc:2023ori}. 
Additionally, in Fig.~\ref{Fig:lowpole_cutoff}, we present the evolution of the lower pole position~\footnote{ 
As to the higher pole, its mass and width have a similar $\Lambda$ dependence.} by varying the momentum cutoff $\Lambda$ from $700$ MeV to $10$ GeV. One can see that the finite-cutoff ($\Lambda=700$ MeV) artifact changes the lower pole position up to $27$ MeV. When $\Lambda$ is up to $3$ GeV, the cutoff-independent result is achieved, which guarantees the renormalizability of the $\pi\Sigma$-$\bar{K}N$ scattering amplitudes in our scheme.

Next, we investigate the evolution of $\Lambda(1405)$ poles by varying the pion mass from $M_\pi\approx 135$ MeV to $400$ MeV along with the quark mass trajectory $\overline{m}=m_\mathrm{symm}$ of the CLS configuration.   

Instead of performing the $\pi\Sigma$-$\bar{K}N$ coupled-channel calculation, we prefer to span all four coupled channels $\pi \Sigma$, $\bar{K} N$, $\eta \Lambda$, and $K \Xi$ to obtain the meson-baryon scattering amplitudes in the $S=-1$ sector. Numerically, the LO potential (Eq.~\eqref{Eq:LOpotential}) varies along with the pion mass since the masses of pseudoscalar mesons, octet baryons, octet vector mesons, and the pion decay constant~\footnote{Note that our description of $F_\pi$ is higher than the lattice data when $M_\pi \leq 200$ MeV, as shown in Fig.~\ref{Fig:MpiMVandFpi}.}, are determined by applying the chiral expansion formulas with the CLS quark mass trajectory $\overline{m}=m_\mathrm{symm}$. Correspondingly, the pion mass dependence also enters the meson-baryon Green function. Thus, by solving the integral equation (Eq.~\eqref{Eq:CCVGT}) and utilizing the subtractive renormalization, one can obtain the renormalized $T$-matrix.  To search poles of $T$-matrix on the complex $s$-plane, one needs to introduce 16 Riemann sheets for the $\pi \Sigma,\, \bar{K} N, \, \eta \Lambda$, and $K \Xi$ couple channels. Similar to the two coupled-channel case, the most relevant Riemann sheets are the physical sheet $(+,+,+,+)$ and the second Riemann sheet $(-,+,+,+)$.

By changing the pion mass from the physical point $M_\pi=135$ MeV to $400$ MeV, we always obtain two poles around the $\pi\Sigma$ and $\bar{K}N$ thresholds in the coupled-channel $T$-matrix. The corresponding pole positions of $\Lambda(1405)$ are presented in Fig.~\ref{Fig:Poles_Mpi} as the function of pion mass. 
The higher pole of $\Lambda(1405)$ is always a resonance, located on the  $(-,+,+,+)$ sheet. Its mass (width) increases (decreases) when the pion mass increases. For the small pion mass region, the higher pole is slightly below the $\bar{K}N$ threshold, while for $M_\pi>225$ MeV, it begins above. This indicates the higher pole couples strongly  to the $\bar{K}N$ channel. Regarding the lower pole of $\Lambda(1405)$, it is intriguing to observe that the evolution of its position becomes more complicated as $M_\pi$ increase, as shown in the right panel of Fig.~\ref{Fig:Poles_Mpi}. At the physical point $M_\pi\approx 135$ MeV, the lower pole is found on the $(-,+,+,+)$ sheet and is recognized as a resonance. Its position $z_R=1320.5-i80.4$ MeV is above the $\pi\Sigma$ threshold with a relatively large imaginary part. As $M_\pi$ increases, the mass of the lower pole slightly rises (but remains below the $\pi\Sigma$ threshold when $M_\pi>146$ MeV), while the width decreases rapidly. Correspondingly, the effective coupling of the lower pole to the $\pi\Sigma$ channel also increases. This indicates that the interaction strength of $\pi\Sigma$ is enhanced along with the increase of $M_\pi$. When the interaction is strong enough, the resonance nature becomes a virtual bound state with $M_\pi\geq 197$ MeV. Because the $\pi\Sigma$ interaction strength is continually enhanced, when $M_\pi\geq 269$ MeV, the lower pole becomes a bound state, which is located on the physical sheet. And the binding energy increases as the pion mass increases. 

From the significantly different pictures of both poles' evaluations, one could infer that the $\bar{K}N$ interaction gradually strengthens with the increase of pion mass. Meanwhile, the strength of the $\pi\Sigma$ interaction changes rapidly and is enhanced sufficiently to render the lower pole of $\Lambda(1405)$ as a bound state. 

We would like to remark that the similar analysis of the evolution of the $\Lambda(1405)$ pole positions at different quark masses has been reported in the $\pi\Sigma$-$\bar{K}N$ coupled-channel study~\cite{Xie:2023cej} by using the Weinberg-Tomozawa term within the traditional CUA, where the light-quark mass dependence of the decay constant and the adjustable subtraction constants are neglected. 
 In contrast, we incorporates the pion mass dependence of all the model variables based on the CLS configuration. Avoid introducing the finite cutoff or the subtractive constants, we obtain the renormalized scattering amplitudes. In this sense that, our LO results of the quark mass dependence of $\Lambda(1405)$, illustrated in Fig.~\ref{Fig:Poles_Mpi}, is a model-independent prediction. The tendency of pole evolution could be checked by the forthcoming studies from the BaSc Collaboration.
  
Furthermore, it is worth mentioning that in Refs.~\cite{Jido:2003cb,Bruns:2021krp,Guo:2023wes}, the evolution of two $\Lambda(1405)$ poles has been studied in the SU(3) trajectory. That's corresponding to a trajectory from the SU(3) flavor limit to the physical point by keeping the average quark mass $\bar{m}=\frac{1}{3}(2m_{u/d}+m_s)$ as a constant and varying a scaling parameter $x\in [0,1]$, which is introduced in the pseudoscalar meson masses, octet baryon masses, the meson decay constants, and the subtraction constants. At LO studies, the higher pole changes from a bound state to a resonance varying $x$ from $0$ to $1$; correspondingly, the lower pole undergoes the bound state to a virtual state and finally comes to a resonance. However, in the recent NLO study~\cite{Guo:2023wes}, the evolutions of both $\Lambda(1405)$ poles along the SU(3) trajectory are changed a lot by including the NLO corrections. It gives us a hint that one needs to go a step further and include the NLO corrections in our framework, at least, to see the changes of the evolution results shown in Fig.~\ref{Fig:Poles_Mpi}. Such a study is beyond the scope of the present manuscript and will be reported in future publications.

\section{Conclusion}~\label{SecIV}
We applied a renormalizable framework of meson-baryon scattering within covariant chiral effective field theory to investigate the $\Lambda(1405)$ resonance in the unphysical quark mass region. Confronting the recent lattice calculation of BaSc Collaboration, our leading order analysis focuses on the quark mass trajectory $\overline{m}=m_\mathrm{symm}$ of the CLS configuration. First, we compared our parameter-free prediction of $\Lambda(1405)$ double poles with the lattice data of BaSc at $M_\pi\approx 200$ MeV and $M_K\approx 487$ MeV. The consistency result indicates the predictive capability of our renormalizable approach. 

Therefore, we investigated the quark mass dependence of the $\Lambda(1405)$ state along with the CLS quark mass trajectory $\overline{m}=m_\mathrm{symm}$. The higher pole of $\Lambda(1405)$ is always a resonance around the $\bar{K}N$ threshold by changing $M_\pi$ from $135$ MeV to $400$ MeV, while the evolution of the lower pole of $\Lambda(1405)$ is complicated: transitioning from a resonance  to a virtual state, and ultimately to a bound state of $\pi\Sigma$ system. Such behavior could be verified by the forthcoming lattice calculation of the BaSc Collaboration.

In order to polish the current results, we plan to extend the study to next-to-leading order (at least) in our renormalizable framework. By incorporating the constraints from the experimental data (see the recent review~\cite{Mai:2020ltx}), such as a large amount of cross sections for $K^- p$ scattering into different final channels, the relevant branching ratios, and the precise energy shift and width of the $1s$ state in kaonic hydrogen by the SIDDHARTA Collaboration~\cite{SIDDHARTA:2011dsy} etc., we can improve the description of the $\bar{K}N$ scattering to better determine the pole positions of $\Lambda(1405)$. At this level, one could provide a moderately reliable prediction about the light-quark dependence of $\Lambda(1405)$ poles. 

\section*{Acknowledgements}
We appreciate the valuable discussions with Daniel Mohler during the MENU 2023 conference. 
This work was supported by the Deutsche Forschungsgemeinschaft (DFG, German Research Foundation), in part through the Research Unit (Photonphoton interactions in the Standard Model and beyond, Projektnummer 458854507—FOR 5327), and in part through the Cluster of Excellence (Precision Physics, Fundamental Interactions, and Structure of Matter) (PRISMA$^+$ EXC 2118/1) within the German Excellence Strategy (Project ID 39083149).

\bibliographystyle{elsarticle-num} 

\bibliography{L1405_mq_depen_PLB_v2}

\begin{thebibliography}{10}
\expandafter\ifx\csname url\endcsname\relax
  \def\url#1{\texttt{#1}}\fi
\expandafter\ifx\csname urlprefix\endcsname\relax\def\urlprefix{URL }\fi
\expandafter\ifx\csname href\endcsname\relax
  \def\href#1#2{#2} \def\path#1{#1}\fi

\bibitem{ParticleDataGroup:2022pth}
R.~L. Workman, et~al., {Review of Particle Physics}, PTEP 2022 (2022) 083C01.
\newblock \href {https://doi.org/10.1093/ptep/ptac097} {\path{doi:10.1093/ptep/ptac097}}.

\bibitem{Barnea:2012qa}
N.~Barnea, A.~Gal, E.~Z. Liverts, {Realistic calculations of $\bar{K} N N$, $\bar{K} N N N$, and $\bar{K} \bar{K} N N$ quasibound states}, Phys. Lett. B 712 (2012) 132--137.
\newblock \href {http://arxiv.org/abs/1203.5234} {\path{arXiv:1203.5234}}, \href {https://doi.org/10.1016/j.physletb.2012.04.055} {\path{doi:10.1016/j.physletb.2012.04.055}}.

\bibitem{Curceanu:2019uph}
C.~Curceanu, et~al., {The modern era of light kaonic atom experiments}, Rev. Mod. Phys. 91~(2) (2019) 025006.
\newblock \href {https://doi.org/10.1103/RevModPhys.91.025006} {\path{doi:10.1103/RevModPhys.91.025006}}.

\bibitem{Kaplan:1986yq}
D.~B. Kaplan, A.~E. Nelson, {Strange Goings on in Dense Nucleonic Matter}, Phys. Lett. B 175 (1986) 57--63.
\newblock \href {https://doi.org/10.1016/0370-2693(86)90331-X} {\path{doi:10.1016/0370-2693(86)90331-X}}.

\bibitem{Li:1997zb}
G.-Q. Li, C.~H. Lee, G.~E. Brown, {Kaons in dense matter, kaon production in heavy ion collisions, and kaon condensation in neutron stars}, Nucl. Phys. A 625 (1997) 372--434.
\newblock \href {http://arxiv.org/abs/nucl-th/9706057} {\path{arXiv:nucl-th/9706057}}, \href {https://doi.org/10.1016/S0375-9474(97)00489-2} {\path{doi:10.1016/S0375-9474(97)00489-2}}.

\bibitem{Gal:2016boi}
A.~Gal, E.~V. Hungerford, D.~J. Millener, {Strangeness in nuclear physics}, Rev. Mod. Phys. 88~(3) (2016) 035004.
\newblock \href {http://arxiv.org/abs/1605.00557} {\path{arXiv:1605.00557}}, \href {https://doi.org/10.1103/RevModPhys.88.035004} {\path{doi:10.1103/RevModPhys.88.035004}}.

\bibitem{Tolos:2020aln}
L.~Tolos, L.~Fabbietti, {Strangeness in Nuclei and Neutron Stars}, Prog. Part. Nucl. Phys. 112 (2020) 103770.
\newblock \href {http://arxiv.org/abs/2002.09223} {\path{arXiv:2002.09223}}, \href {https://doi.org/10.1016/j.ppnp.2020.103770} {\path{doi:10.1016/j.ppnp.2020.103770}}.

\bibitem{Dalitz:1959dn}
R.~H. Dalitz, S.~F. Tuan, {A possible resonant state in pion-hyperon scattering}, Phys. Rev. Lett. 2 (1959) 425--428.
\newblock \href {https://doi.org/10.1103/PhysRevLett.2.425} {\path{doi:10.1103/PhysRevLett.2.425}}.

\bibitem{Dalitz:1960du}
R.~H. Dalitz, S.~F. Tuan, {The phenomenological description of -K -nucleon reaction processes}, Annals Phys. 10 (1960) 307--351.
\newblock \href {https://doi.org/10.1016/0003-4916(60)90001-4} {\path{doi:10.1016/0003-4916(60)90001-4}}.

\bibitem{Alston:1961zzd}
M.~H. Alston, L.~W. Alvarez, P.~Eberhard, M.~L. Good, W.~Graziano, H.~K. Ticho, S.~G. Wojcicki, {Study of Resonances of the Sigma-pi System}, Phys. Rev. Lett. 6 (1961) 698--702.
\newblock \href {https://doi.org/10.1103/PhysRevLett.6.698} {\path{doi:10.1103/PhysRevLett.6.698}}.

\bibitem{Isgur:1978xj}
N.~Isgur, G.~Karl, {P Wave Baryons in the Quark Model}, Phys. Rev. D 18 (1978) 4187.
\newblock \href {https://doi.org/10.1103/PhysRevD.18.4187} {\path{doi:10.1103/PhysRevD.18.4187}}.

\bibitem{Weinberg:1978kz}
S.~Weinberg, {Phenomenological Lagrangians}, Physica A 96~(1-2) (1979) 327--340.
\newblock \href {https://doi.org/10.1016/0378-4371(79)90223-1} {\path{doi:10.1016/0378-4371(79)90223-1}}.

\bibitem{Oller:2000ma}
J.~A. Oller, E.~Oset, A.~Ramos, {Chiral unitary approach to meson meson and meson - baryon interactions and nuclear applications}, Prog. Part. Nucl. Phys. 45 (2000) 157--242.
\newblock \href {http://arxiv.org/abs/hep-ph/0002193} {\path{arXiv:hep-ph/0002193}}, \href {https://doi.org/10.1016/S0146-6410(00)00104-6} {\path{doi:10.1016/S0146-6410(00)00104-6}}.

\bibitem{Oller:2000fj}
J.~A. Oller, U.~G. Mei\ss{}ner, {Chiral dynamics in the presence of bound states: Kaon nucleon interactions revisited}, Phys. Lett. B 500 (2001) 263--272.
\newblock \href {http://arxiv.org/abs/hep-ph/0011146} {\path{arXiv:hep-ph/0011146}}, \href {https://doi.org/10.1016/S0370-2693(01)00078-8} {\path{doi:10.1016/S0370-2693(01)00078-8}}.

\bibitem{Mai:2020ltx}
M.~Mai, {Review of the ${\Lambda }$(1405) A curious case of a strangeness resonance}, Eur. Phys. J. ST 230~(6) (2021) 1593--1607.
\newblock \href {http://arxiv.org/abs/2010.00056} {\path{arXiv:2010.00056}}, \href {https://doi.org/10.1140/epjs/s11734-021-00144-7} {\path{doi:10.1140/epjs/s11734-021-00144-7}}.

\bibitem{Hyodo:2020czb}
T.~Hyodo, M.~Niiyama, {QCD and the strange baryon spectrum}, Prog. Part. Nucl. Phys. 120 (2021) 103868.
\newblock \href {http://arxiv.org/abs/2010.07592} {\path{arXiv:2010.07592}}, \href {https://doi.org/10.1016/j.ppnp.2021.103868} {\path{doi:10.1016/j.ppnp.2021.103868}}.

\bibitem{Meissner:2020khl}
U.-G. Mei\ss{}ner, {Two-pole structures in QCD: Facts, not fantasy!}, Symmetry 12~(6) (2020) 981.
\newblock \href {http://arxiv.org/abs/2005.06909} {\path{arXiv:2005.06909}}, \href {https://doi.org/10.3390/sym12060981} {\path{doi:10.3390/sym12060981}}.

\bibitem{BGOOD:2021sog}
G.~Scheluchin, et~al., {Photoproduction of K+\ensuremath{\Lambda}(1405)\textrightarrow{}K+\ensuremath{\pi}0\ensuremath{\Sigma}0 extending to forward angles and low momentum transfer}, Phys. Lett. B 833 (2022) 137375.
\newblock \href {http://arxiv.org/abs/2108.12235} {\path{arXiv:2108.12235}}, \href {https://doi.org/10.1016/j.physletb.2022.137375} {\path{doi:10.1016/j.physletb.2022.137375}}.

\bibitem{Piscicchia:2022wmd}
K.~Piscicchia, et~al., {First simultaneous K\ensuremath{-}p\textrightarrow{}\ensuremath{\Sigma}0\ensuremath{\pi}0,~\ensuremath{\Lambda}\ensuremath{\pi}0 cross section~measurements at 98 MeV/c}, Phys. Rev. C 108~(5) (2023) 055201.
\newblock \href {http://arxiv.org/abs/2210.10342} {\path{arXiv:2210.10342}}, \href {https://doi.org/10.1103/PhysRevC.108.055201} {\path{doi:10.1103/PhysRevC.108.055201}}.

\bibitem{J-PARCE31:2022plu}
S.~Aikawa, et~al., {Pole position of \ensuremath{\Lambda}(1405) measured in d(K\ensuremath{-},n)\ensuremath{\pi}\ensuremath{\Sigma} reactions}, Phys. Lett. B 837 (2023) 137637.
\newblock \href {http://arxiv.org/abs/2209.08254} {\path{arXiv:2209.08254}}, \href {https://doi.org/10.1016/j.physletb.2022.137637} {\path{doi:10.1016/j.physletb.2022.137637}}.

\bibitem{ALICE:2022yyh}
S.~Acharya, et~al., {Constraining the ${\overline{\textrm{K}}}{\textrm{N}}$ coupled channel dynamics using femtoscopic correlations at the LHC}, Eur. Phys. J. C 83~(4) (2023) 340.
\newblock \href {http://arxiv.org/abs/2205.15176} {\path{arXiv:2205.15176}}, \href {https://doi.org/10.1140/epjc/s10052-023-11476-0} {\path{doi:10.1140/epjc/s10052-023-11476-0}}.

\bibitem{Lu:2022hwm}
J.-X. Lu, L.-S. Geng, M.~Doering, M.~Mai, {Cross-Channel Constraints on Resonant Antikaon-Nucleon Scattering}, Phys. Rev. Lett. 130~(7) (2023) 071902.
\newblock \href {http://arxiv.org/abs/2209.02471} {\path{arXiv:2209.02471}}, \href {https://doi.org/10.1103/PhysRevLett.130.071902} {\path{doi:10.1103/PhysRevLett.130.071902}}.

\bibitem{Sadasivan:2022srs}
D.~Sadasivan, M.~Mai, M.~D\"oring, U.-G. Mei\ss{}ner, F.~Amorim, J.~P. Klucik, J.-X. Lu, L.-S. Geng, {New insights into the pole parameters of the $\Lambda(1380)$, the $\Lambda(1405)$ and the $\Sigma(1385)$}, Front. Phys. 11 (2023) 1139236.
\newblock \href {http://arxiv.org/abs/2212.10415} {\path{arXiv:2212.10415}}, \href {https://doi.org/10.3389/fphy.2023.1139236} {\path{doi:10.3389/fphy.2023.1139236}}.

\bibitem{Cieply:2023saa}
A.~Cieply, P.~C. Bruns, {Constraining the chirally motivated \ensuremath{\pi}\ensuremath{\Sigma}\ensuremath{-}K\textasciimacron{}N models with the \ensuremath{\pi}\ensuremath{\Sigma} photoproduction mass spectra}, Nucl. Phys. A 1043 (2024) 122819.
\newblock \href {http://arxiv.org/abs/2305.06205} {\path{arXiv:2305.06205}}, \href {https://doi.org/10.1016/j.nuclphysa.2024.122819} {\path{doi:10.1016/j.nuclphysa.2024.122819}}.

\bibitem{Nemoto:2003ft}
Y.~Nemoto, N.~Nakajima, H.~Matsufuru, H.~Suganuma, {Negative parity baryons in quenched anisotropic lattice QCD}, Phys. Rev. D 68 (2003) 094505.
\newblock \href {http://arxiv.org/abs/hep-lat/0302013} {\path{arXiv:hep-lat/0302013}}, \href {https://doi.org/10.1103/PhysRevD.68.094505} {\path{doi:10.1103/PhysRevD.68.094505}}.

\bibitem{Menadue:2011pd}
B.~J. Menadue, W.~Kamleh, D.~B. Leinweber, M.~S. Mahbub, {Isolating the $\Lambda(1405)$ in Lattice QCD}, Phys. Rev. Lett. 108 (2012) 112001.
\newblock \href {http://arxiv.org/abs/1109.6716} {\path{arXiv:1109.6716}}, \href {https://doi.org/10.1103/PhysRevLett.108.112001} {\path{doi:10.1103/PhysRevLett.108.112001}}.

\bibitem{Engel:2013ig}
G.~P. Engel, C.~B. Lang, D.~Mohler, A.~Sch\"afer, {QCD with Two Light Dynamical Chirally Improved Quarks: Baryons}, Phys. Rev. D 87~(7) (2013) 074504.
\newblock \href {http://arxiv.org/abs/1301.4318} {\path{arXiv:1301.4318}}, \href {https://doi.org/10.1103/PhysRevD.87.074504} {\path{doi:10.1103/PhysRevD.87.074504}}.

\bibitem{Hall:2014uca}
J.~M.~M. Hall, W.~Kamleh, D.~B. Leinweber, B.~J. Menadue, B.~J. Owen, A.~W. Thomas, R.~D. Young, {Lattice QCD Evidence that the \ensuremath{\Lambda}(1405) Resonance is an Antikaon-Nucleon Molecule}, Phys. Rev. Lett. 114~(13) (2015) 132002.
\newblock \href {http://arxiv.org/abs/1411.3402} {\path{arXiv:1411.3402}}, \href {https://doi.org/10.1103/PhysRevLett.114.132002} {\path{doi:10.1103/PhysRevLett.114.132002}}.

\bibitem{Meinel:2021grq}
S.~Meinel, G.~Rendon, {Charm-baryon semileptonic decays and the strange \ensuremath{\Lambda}* resonances: New insights from lattice QCD}, Phys. Rev. D 105~(5) (2022) L051505.
\newblock \href {http://arxiv.org/abs/2107.13084} {\path{arXiv:2107.13084}}, \href {https://doi.org/10.1103/PhysRevD.105.L051505} {\path{doi:10.1103/PhysRevD.105.L051505}}.

\bibitem{BaryonScatteringBaSc:2023zvt}
J.~Bulava, et~al., {Two-Pole Nature of the \ensuremath{\Lambda}(1405) resonance from Lattice QCD}, Phys. Rev. Lett. 132~(5) (2024) 051901.
\newblock \href {http://arxiv.org/abs/2307.10413} {\path{arXiv:2307.10413}}, \href {https://doi.org/10.1103/PhysRevLett.132.051901} {\path{doi:10.1103/PhysRevLett.132.051901}}.

\bibitem{BaryonScatteringBaSc:2023ori}
J.~Bulava, et~al., {Lattice QCD study of \ensuremath{\pi}\ensuremath{\Sigma}-K\textasciimacron{}N scattering and the \ensuremath{\Lambda}(1405) resonance}, Phys. Rev. D 109~(1) (2024) 014511.
\newblock \href {http://arxiv.org/abs/2307.13471} {\path{arXiv:2307.13471}}, \href {https://doi.org/10.1103/PhysRevD.109.014511} {\path{doi:10.1103/PhysRevD.109.014511}}.

\bibitem{Ren:2020wid}
X.-L. Ren, E.~Epelbaum, J.~Gegelia, U.-G. Mei\ss{}ner, {Meson\textendash{}baryon scattering in resummed baryon chiral perturbation theory using time-ordered perturbation theory}, Eur. Phys. J. C 80~(5) (2020) 406.
\newblock \href {http://arxiv.org/abs/2003.06272} {\path{arXiv:2003.06272}}, \href {https://doi.org/10.1140/epjc/s10052-020-7991-x} {\path{doi:10.1140/epjc/s10052-020-7991-x}}.

\bibitem{Epelbaum:2020maf}
E.~Epelbaum, A.~M. Gasparyan, J.~Gegelia, U.-G. Mei\ss{}ner, X.-L. Ren, {How to renormalize integral equations with singular potentials in effective field theory}, Eur. Phys. J. A 56~(5) (2020) 152.
\newblock \href {http://arxiv.org/abs/2001.07040} {\path{arXiv:2001.07040}}, \href {https://doi.org/10.1140/epja/s10050-020-00162-4} {\path{doi:10.1140/epja/s10050-020-00162-4}}.

\bibitem{Ren:2021yxc}
X.-L. Ren, E.~Epelbaum, J.~Gegelia, U.-G. Mei\ss{}ner, {The $\Lambda (1405)$ in resummed chiral effective field theory}, Eur. Phys. J. C 81~(7) (2021) 582.
\newblock \href {http://arxiv.org/abs/2102.00914} {\path{arXiv:2102.00914}}, \href {https://doi.org/10.1140/epjc/s10052-021-09386-0} {\path{doi:10.1140/epjc/s10052-021-09386-0}}.

\bibitem{Mai:2014xna}
M.~Mai, U.-G. Mei\ss{}ner, {Constraints on the chiral unitary $\bar KN$ amplitude from $\pi\Sigma K^+$ photoproduction data}, Eur. Phys. J. A 51~(3) (2015) 30.
\newblock \href {http://arxiv.org/abs/1411.7884} {\path{arXiv:1411.7884}}, \href {https://doi.org/10.1140/epja/i2015-15030-3} {\path{doi:10.1140/epja/i2015-15030-3}}.

\bibitem{Jido:2003cb}
D.~Jido, J.~A. Oller, E.~Oset, A.~Ramos, U.~G. Mei\ss{}ner, {Chiral dynamics of the two Lambda(1405) states}, Nucl. Phys. A 725 (2003) 181--200.
\newblock \href {http://arxiv.org/abs/nucl-th/0303062} {\path{arXiv:nucl-th/0303062}}, \href {https://doi.org/10.1016/S0375-9474(03)01598-7} {\path{doi:10.1016/S0375-9474(03)01598-7}}.

\bibitem{Garcia-Recio:2003ejq}
C.~Garcia-Recio, M.~F.~M. Lutz, J.~Nieves, {Quark mass dependence of s wave baryon resonances}, Phys. Lett. B 582 (2004) 49--54.
\newblock \href {http://arxiv.org/abs/nucl-th/0305100} {\path{arXiv:nucl-th/0305100}}, \href {https://doi.org/10.1016/j.physletb.2003.11.073} {\path{doi:10.1016/j.physletb.2003.11.073}}.

\bibitem{Molina:2015uqp}
R.~Molina, M.~D\"oring, {Pole structure of the $\Lambda$(1405) in a recent QCD simulation}, Phys. Rev. D 94~(5) (2016) 056010, [Addendum: Phys.Rev.D 94, 079901 (2016)].
\newblock \href {http://arxiv.org/abs/1512.05831} {\path{arXiv:1512.05831}}, \href {https://doi.org/10.1103/PhysRevD.94.079901} {\path{doi:10.1103/PhysRevD.94.079901}}.

\bibitem{Bruns:2021krp}
P.~C. Bruns, A.~Ciepl\'y, {SU(3) flavor symmetry considerations for the K\textasciimacron{}N coupled channels system}, Nucl. Phys. A 1019 (2022) 122378.
\newblock \href {http://arxiv.org/abs/2109.03109} {\path{arXiv:2109.03109}}, \href {https://doi.org/10.1016/j.nuclphysa.2021.122378} {\path{doi:10.1016/j.nuclphysa.2021.122378}}.

\bibitem{Xie:2023cej}
J.-M. Xie, J.-X. Lu, L.-S. Geng, B.-S. Zou, {Two-pole structures as a universal phenomenon dictated by coupled-channel chiral dynamics}, Phys. Rev. D 108~(11) (2023) L111502.
\newblock \href {http://arxiv.org/abs/2307.11631} {\path{arXiv:2307.11631}}, \href {https://doi.org/10.1103/PhysRevD.108.L111502} {\path{doi:10.1103/PhysRevD.108.L111502}}.

\bibitem{Guo:2023wes}
F.-K. Guo, Y.~Kamiya, M.~Mai, U.-G. Mei\ss{}ner, {New insights into the nature of the \ensuremath{\Lambda}(1380) and \ensuremath{\Lambda}(1405) resonances away from the SU(3) limit}, Phys. Lett. B 846 (2023) 138264.
\newblock \href {http://arxiv.org/abs/2308.07658} {\path{arXiv:2308.07658}}, \href {https://doi.org/10.1016/j.physletb.2023.138264} {\path{doi:10.1016/j.physletb.2023.138264}}.

\bibitem{RQCD:2022xux}
G.~S. Bali, S.~Collins, P.~Georg, D.~Jenkins, P.~Korcyl, A.~Sch\"afer, E.~E. Scholz, J.~Simeth, W.~S\"oldner, S.~Weish\"aupl, {Scale setting and the light baryon spectrum in N$_{f}$ = 2 + 1 QCD with Wilson fermions}, JHEP 05 (2023) 035.
\newblock \href {http://arxiv.org/abs/2211.03744} {\path{arXiv:2211.03744}}, \href {https://doi.org/10.1007/JHEP05(2023)035} {\path{doi:10.1007/JHEP05(2023)035}}.

\bibitem{Ce:2022eix}
M.~C\`e, A.~G\'erardin, G.~von Hippel, H.~B. Meyer, K.~Miura, K.~Ottnad, A.~Risch, T.~San~Jos\'e, J.~Wilhelm, H.~Wittig, {The hadronic running of the electromagnetic coupling and the electroweak mixing angle from lattice QCD}, JHEP 08 (2022) 220.
\newblock \href {http://arxiv.org/abs/2203.08676} {\path{arXiv:2203.08676}}, \href {https://doi.org/10.1007/JHEP08(2022)220} {\path{doi:10.1007/JHEP08(2022)220}}.

\bibitem{Ce:2022kxy}
M.~C\`e, et~al., {Window observable for the hadronic vacuum polarization contribution to the muon g-2 from lattice QCD}, Phys. Rev. D 106~(11) (2022) 114502.
\newblock \href {http://arxiv.org/abs/2206.06582} {\path{arXiv:2206.06582}}, \href {https://doi.org/10.1103/PhysRevD.106.114502} {\path{doi:10.1103/PhysRevD.106.114502}}.

\bibitem{Sadasivan:2021emk}
D.~Sadasivan, A.~Alexandru, H.~Akdag, F.~Amorim, R.~Brett, C.~Culver, M.~D\"oring, F.~X. Lee, M.~Mai, {Pole position of the a1(1260) resonance in a three-body unitary framework}, Phys. Rev. D 105~(5) (2022) 054020.
\newblock \href {http://arxiv.org/abs/2112.03355} {\path{arXiv:2112.03355}}, \href {https://doi.org/10.1103/PhysRevD.105.054020} {\path{doi:10.1103/PhysRevD.105.054020}}.

\bibitem{Draper:2023xvu}
Z.~T. Draper, M.~T. Hansen, F.~Romero-L\'opez, S.~R. Sharpe, {Three relativistic neutrons in a finite volume}, JHEP 07 (2023) 226.
\newblock \href {http://arxiv.org/abs/2303.10219} {\path{arXiv:2303.10219}}, \href {https://doi.org/10.1007/JHEP07(2023)226} {\path{doi:10.1007/JHEP07(2023)226}}.

\bibitem{Kawarabayashi:1966kd}
K.~Kawarabayashi, M.~Suzuki, {Partially conserved axial vector current and the decays of vector mesons}, Phys. Rev. Lett. 16 (1966) 255.
\newblock \href {https://doi.org/10.1103/PhysRevLett.16.255} {\path{doi:10.1103/PhysRevLett.16.255}}.

\bibitem{Riazuddin:1966sw}
Riazuddin, Fayyazuddin, {Algebra of current components and decay widths of rho and K* mesons}, Phys. Rev. 147 (1966) 1071--1073.
\newblock \href {https://doi.org/10.1103/PhysRev.147.1071} {\path{doi:10.1103/PhysRev.147.1071}}.

\bibitem{Djukanovic:2004mm}
D.~Djukanovic, M.~R. Schindler, J.~Gegelia, G.~Japaridze, S.~Scherer, {Universality of the rho-meson coupling in effective field theory}, Phys. Rev. Lett. 93 (2004) 122002.
\newblock \href {http://arxiv.org/abs/hep-ph/0407239} {\path{arXiv:hep-ph/0407239}}, \href {https://doi.org/10.1103/PhysRevLett.93.122002} {\path{doi:10.1103/PhysRevLett.93.122002}}.

\bibitem{Baru:2019ndr}
V.~Baru, E.~Epelbaum, J.~Gegelia, X.-L. Ren, {Towards baryon-baryon scattering in manifestly Lorentz-invariant formulation of SU(3) baryon chiral perturbation theory}, Phys. Lett. B 798 (2019) 134987.
\newblock \href {http://arxiv.org/abs/1905.02116} {\path{arXiv:1905.02116}}, \href {https://doi.org/10.1016/j.physletb.2019.134987} {\path{doi:10.1016/j.physletb.2019.134987}}.

\bibitem{Ren:2019qow}
X.-L. Ren, E.~Epelbaum, J.~Gegelia, {$\Lambda$ -nucleon scattering in baryon chiral perturbation theory}, Phys. Rev. C 101~(3) (2020) 034001.
\newblock \href {http://arxiv.org/abs/1911.05616} {\path{arXiv:1911.05616}}, \href {https://doi.org/10.1103/PhysRevC.101.034001} {\path{doi:10.1103/PhysRevC.101.034001}}.

\bibitem{Fuchs:2003qc}
T.~Fuchs, J.~Gegelia, G.~Japaridze, S.~Scherer, {Renormalization of relativistic baryon chiral perturbation theory and power counting}, Phys. Rev. D 68 (2003) 056005.
\newblock \href {http://arxiv.org/abs/hep-ph/0302117} {\path{arXiv:hep-ph/0302117}}, \href {https://doi.org/10.1103/PhysRevD.68.056005} {\path{doi:10.1103/PhysRevD.68.056005}}.

\bibitem{Ren:2012aj}
X.-L. Ren, L.~S. Geng, J.~Martin~Camalich, J.~Meng, H.~Toki, {Octet baryon masses in next-to-next-to-next-to-leading order covariant baryon chiral perturbation theory}, JHEP 12 (2012) 073.
\newblock \href {http://arxiv.org/abs/1209.3641} {\path{arXiv:1209.3641}}, \href {https://doi.org/10.1007/JHEP12(2012)073} {\path{doi:10.1007/JHEP12(2012)073}}.

\bibitem{Aoki:2016frl}
S.~Aoki, et~al., {Review of lattice results concerning low-energy particle physics}, Eur. Phys. J. C 77~(2) (2017) 112.
\newblock \href {http://arxiv.org/abs/1607.00299} {\path{arXiv:1607.00299}}, \href {https://doi.org/10.1140/epjc/s10052-016-4509-7} {\path{doi:10.1140/epjc/s10052-016-4509-7}}.

\bibitem{Zhou:2014ila}
Y.~Zhou, X.-L. Ren, H.-X. Chen, L.-S. Geng, {Pseudoscalar meson and vector meson interactions and dynamically generated axial-vector mesons}, Phys. Rev. D 90~(1) (2014) 014020.
\newblock \href {http://arxiv.org/abs/1404.6847} {\path{arXiv:1404.6847}}, \href {https://doi.org/10.1103/PhysRevD.90.014020} {\path{doi:10.1103/PhysRevD.90.014020}}.

\bibitem{Nebreda:2010wv}
J.~Nebreda, J.~R. Pelaez., {Strange and non-strange quark mass dependence of elastic light resonances from SU(3) Unitarized Chiral Perturbation Theory to one loop}, Phys. Rev. D 81 (2010) 054035.
\newblock \href {http://arxiv.org/abs/1001.5237} {\path{arXiv:1001.5237}}, \href {https://doi.org/10.1103/PhysRevD.81.054035} {\path{doi:10.1103/PhysRevD.81.054035}}.

\bibitem{Epelbaum:2008ga}
E.~Epelbaum, H.-W. Hammer, U.-G. Mei\ss{}ner, {Modern Theory of Nuclear Forces}, Rev. Mod. Phys. 81 (2009) 1773--1825.
\newblock \href {http://arxiv.org/abs/0811.1338} {\path{arXiv:0811.1338}}, \href {https://doi.org/10.1103/RevModPhys.81.1773} {\path{doi:10.1103/RevModPhys.81.1773}}.

\bibitem{Machleidt:2011zz}
R.~Machleidt, D.~R. Entem, {Chiral effective field theory and nuclear forces}, Phys. Rept. 503 (2011) 1--75.
\newblock \href {http://arxiv.org/abs/1105.2919} {\path{arXiv:1105.2919}}, \href {https://doi.org/10.1016/j.physrep.2011.02.001} {\path{doi:10.1016/j.physrep.2011.02.001}}.

\bibitem{Bruns:2010sv}
P.~C. Bruns, M.~Mai, U.~G. Mei\ss{}ner, {Chiral dynamics of the S11(1535) and S11(1650) resonances revisited}, Phys. Lett. B 697 (2011) 254--259.
\newblock \href {http://arxiv.org/abs/1012.2233} {\path{arXiv:1012.2233}}, \href {https://doi.org/10.1016/j.physletb.2011.02.008} {\path{doi:10.1016/j.physletb.2011.02.008}}.

\bibitem{Mai:2012dt}
M.~Mai, U.-G. Mei\ss{}ner, {New insights into antikaon-nucleon scattering and the structure of the Lambda(1405)}, Nucl. Phys. A 900 (2013) 51 -- 64.
\newblock \href {http://arxiv.org/abs/1202.2030} {\path{arXiv:1202.2030}}, \href {https://doi.org/10.1016/j.nuclphysa.2013.01.032} {\path{doi:10.1016/j.nuclphysa.2013.01.032}}.

\bibitem{Morimatsu:2019wvk}
O.~Morimatsu, K.~Yamada, {Renormalization of the unitarized Weinberg-Tomozawa interaction without on-shell factorization and I=0 K\textasciimacron{}N\textendash{}\ensuremath{\pi}\ensuremath{\Sigma} coupled channels}, Phys. Rev. C 100~(2) (2019) 025201.
\newblock \href {http://arxiv.org/abs/1903.12380} {\path{arXiv:1903.12380}}, \href {https://doi.org/10.1103/PhysRevC.100.025201} {\path{doi:10.1103/PhysRevC.100.025201}}.

\bibitem{SIDDHARTA:2011dsy}
M.~Bazzi, et~al., {A New Measurement of Kaonic Hydrogen X-rays}, Phys. Lett. B 704 (2011) 113--117.
\newblock \href {http://arxiv.org/abs/1105.3090} {\path{arXiv:1105.3090}}, \href {https://doi.org/10.1016/j.physletb.2011.09.011} {\path{doi:10.1016/j.physletb.2011.09.011}}.

\end{thebibliography}

\end{document}